\documentclass[useAMS,usenatbib,fleqn]{mnras}
\usepackage{graphicx,subfig,lscape}
\usepackage[usenames,dvipsnames]{xcolor}
\usepackage{times}

\newcommand{\kms}{km\,s$^{-1}$}
\newcommand{\vs}{$v \sin i$}

\newcommand{\teff}{$T_{\rm eff}$}
\newcommand{\lgg}{$\log\,{g}$}

\title[The magnetic field of 49 Cam]{The complex magnetic field topology of the cool Ap star 49 Cam}

\author[J. Silvester et al.]
{J. Silvester$^{1}$, O.Kochukhov$^{1}$, N.Rusomarov$^{1}$ and G.A. Wade$^{2}$\\
$^{1}$Department of Astronomy and Space Physics, Uppsala University, 751 20, Uppsala, Sweden\\
$^{2}$Department of Physics, Royal Military College of Canada, P.O. Box 17000, Station `Forces', Kingston, Ontario, Canada, K7K 7B4\\
}  
\begin{document}

\date{Accepted . Received }

\pagerange{\pageref{firstpage}--\pageref{lastpage}} \pubyear{2017}

\maketitle

\label{firstpage}

\begin{abstract}
49 Cam is a cool magnetic chemically peculiar star which has been noted for showing strong, complex Zeeman linear polarisation signatures.  This paper describes magnetic and chemical surface maps obtained for 49 Cam using the {\sc Invers10} magnetic Doppler imaging code and high-resolution spectropolarimetric data in all four Stokes parameters collected with the ESPaDOnS and Narval spectropolarimeters at the Canada-France-Hawaii Telescope and Pic du Midi Observatory. The reconstructed magnetic field maps of 49 Cam show a relatively complex structure. Describing the magnetic field topology in terms of spherical harmonics, we find significant contributions of modes up to $\ell=3$, including toroidal components. Observations cannot be reproduced using a simple low-order multipolar magnetic field structure. 49 Cam exhibits a level of field complexity that has not been seen in magnetic maps of other cool Ap stars. Hence we concluded that relatively complex magnetic fields are observed in Ap stars at both low and high effective temperatures. In addition to mapping the magnetic field, we also derive surface abundance distributions of nine chemical elements, including Ca, Sc, Ti, Cr, Fe, Ce, Pr, Nd, Eu. Comparing these abundance maps with the reconstructed magnetic field geometry, we find no clear relationship of the abundance distributions with the magnetic field for some elements. However, for other elements some distinct patterns are found. We discuss these results in the context of other recent magnetic mapping studies and theoretical predictions of radiative diffusion. 
\end{abstract}


\section{Introduction}

The intermediate-mass Ap/Bp stars are A and B-type main sequence stars which display peculiar surface abundance patterns when compared to solar abundances. These abundance peculiarities are typically associated with a strong, global and stable magnetic field. These fields have strengths ranging from hundreds of G up to tens of kG. Historically, the magnetic topologies of Ap/Bp stars have been considered to be close to a simple dipolar structure, but more recently it has been shown that the surface fields of some stars also contain significant contributions from complex substructures superimposed on a strong quasi-dipolar component \citep[e.g.][]{Kochukhov10}. The observed peculiar abundances and abundance structures are thought to be a result of the build-up of chemical stratification in the radiative stellar envelope under the influence of radiative levitation and gravitational settling, mediated by the magnetic field and mixing processes \citep{Michaud15}.

The standard phenomenological model for describing the variability of Ap/Bp stars is the oblique rotator model \citep{Stibbs:1950aa}, which describes a rotating star whose magnetic axis is inclined relative to its rotational axis.  By exploiting the rotation of these objects, Ap/Bp stars are often studied using Doppler imaging techniques for the purpose of reconstructing a two-dimensional surface maps of chemical and magnetic field structures using spectroscopic or spectropolarimetric time-series observations \citep{Kochukhov16}.  

The magnetic field geometry and the geometry of the chemical surface inhomogeneities at the surfaces of Ap/Bp stars gives a key insight into the physical mechanisms governing the global magnetic field evolution in stellar interiors \citep{Braithwaite:2006aa,Duez:2010aa}. By studying the magnetic field and chemical spots in detail, we can obtain empirical constraints on both the geometry of the field for theoretical modelling of the magnetised stellar interiors \citep[e.g.][]{Braithwaite:2006aa,Duez:2010ab} and for testing theories of atomic diffusion, thought to be responsible for the formation of chemical abundance inhomogeneities \citep[e.g.][]{Alecian15}.

The Ap star 49 Cam (HD 62140) was classified as F0p (SrEu) by \citet{Cowley68} and later as A8p (SrEu) by \citet{Leone2000}.  An early detailed study of 49 Cam was carried out by \citet{Bonsack74}, who examined the photometric and spectroscopic variations of this star.  Later, a detailed analysis of the magnetic field geometry was performed by \citet{Leroy94} and \citet{Leroy96} using broadband linear polarisation measurements. 49 Cam was found to exhibit very strong linear polarisation signatures by \citet{Leroy94}, who also demonstrated that the surface magnetic field of this star could not be described with a simple dipolar configuration. This work was an early indication of the complex field structure of 49 Cam, and is a principle reason for including this star in our magnetic mapping programme. 

High resolution spectropolarimetric observations of 49 Cam have been obtained in Stokes $IQUV$ on two occasions: first by \citet{Wade2000} and then by \citet{Silvester12} as part of the dataset obtained with the ESPaDOnS and Narval spectropolarimeters. In this paper we use the Stokes parameter spectra collected by \citet{Silvester12} to derive the magnetic and chemical maps of 49 Cam. This spectropolarimetric dataset and similar datasets from the HARPSpol instrument at the ESO 3.6-m telescope have been successfully used for mapping the magnetic fields of a growing number of Ap stars, e.g. HD 32633 \citep{Silvester15}, $\alpha^2$\,CVn \citep{Silvester14a,Silvester14b}, HD 24712 \citep{Rusomarov15} and HD 125248 \citep{Rusomarov16}. These magnetic mapping studies have relied on using the magnetic Doppler imaging (MDI) inversion code {\sc Invers10} \citep{Piskunov02,Kochukhov02}, which models four Stokes parameters line profiles with detailed polarised radiative transfer calculations and can simultaneously reconstruct the surface vector magnetic field maps and chemical abundance distributions. These MDI studies indicate that many Ap/Bp stars when modelling using 4 Stokes parameter datasets have magnetic fields which deviate from a low-order multipolar configuration and, especially, cannot be represented as oblique dipoles. For a complete list of Ap/Bp stars studied with the Doppler or magnetic Doppler imaging techniques we refer to \citet{Kochukhov17}; this paper also provides a series of tests of the reliability of chemical abundance maps reconstructed with {\sc Invers10}.

The paper is organised in the following way. Section~\ref{sect:obs} describes briefly the observational data. In Section~\ref{sect:params} the derivation of stellar parameters is discussed. Section~\ref{sect:methods} the procedure for magnetic Doppler imaging is described.  In Sections~\ref{sect:mag} and \ref{sect:abn} the final magnetic field map and chemical abundance maps are presented respectively. The results and implications of this work are discussed in Section~\ref{sect:discuss}.

\section{Spectropolarimetric observations}
\label{sect:obs}
Observations of 49 Cam analysed in this paper were obtained between 2006 and 2010, with both the ESPaDOnS and Narval spectropolarimeters. A detailed discussion of these observations, instruments, observing procedures, and all essential aspects of data reduction is provided by \citet{Silvester12}. In total, 19 Stokes $IQUV$ observations of 49 Cam are available. The spectra have a resolving power of 65,000, cover the wavelength range of 369--1048~nm, and have a typical signal-to-noise ratio of 600:1.  

\begin{figure*}
\begin{center}
 \includegraphics[width=0.94\textwidth]{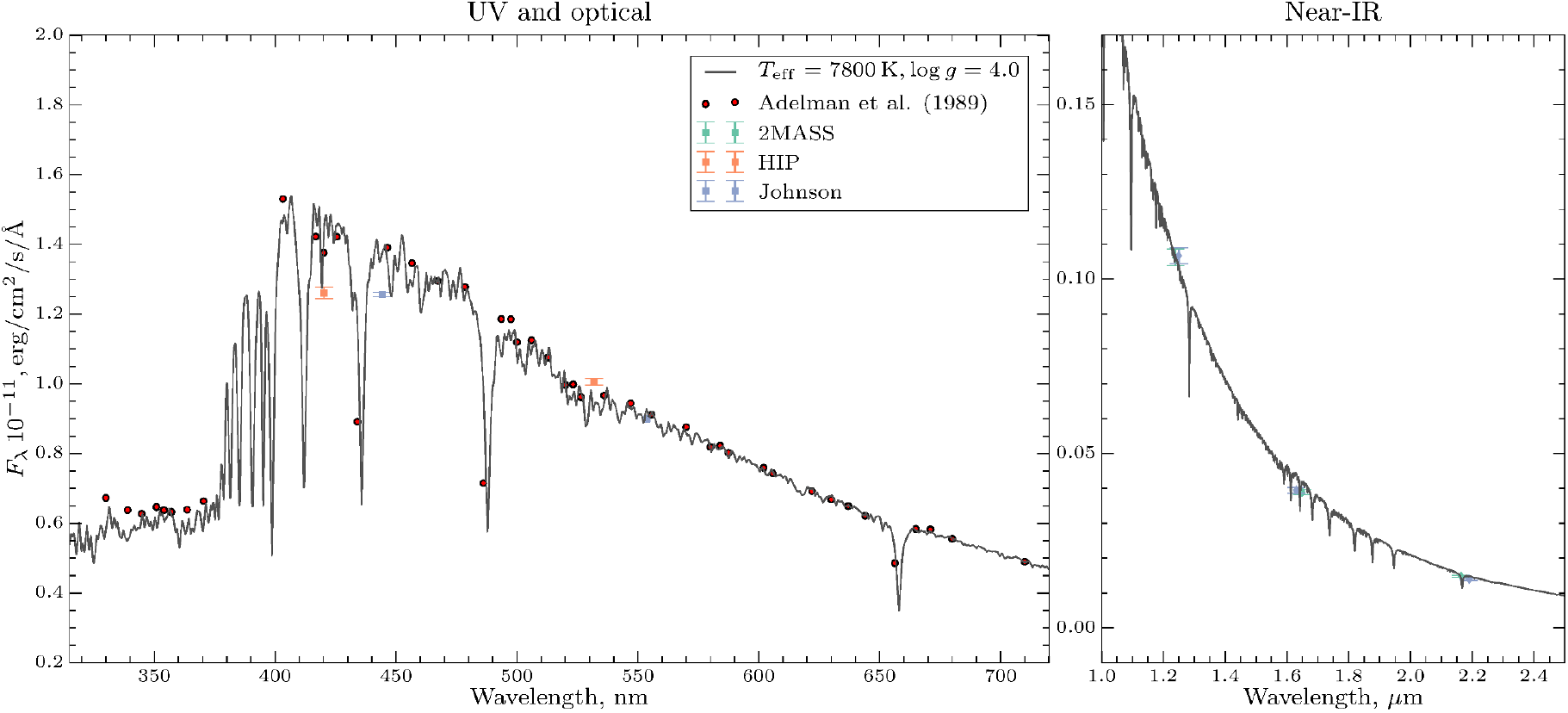}
  \caption{Comparison between theoretical spectral energy distribution (solid line) and observations (symbols) in the near-UV, optical and near-IR spectral regions. The theoretical SED corresponds to the calculations for $T_{\rm eff}=7800$ K and $\log g=4.0$, including the effects of individual non-solar abundances and magnetic field.}
\label{SED}
\end{center}
\end{figure*}

\section{Stellar Parameters}
\label{sect:params}

49~Cam is a fairly broad-lined Ap star with an effective temperature in the range 7700--8000~K \citep{Adel82,Babel94,Ryabchikova01}. We re-determined the stellar effective temperature by computing a small grid (with steps of 50K) of model atmospheres within the range of $T_{\rm eff}$ reported in the literature using the {\sc LLmodels} code \citep{Shulyak04} and comparing the resulting theoretical spectral energy distributions (SED) with the observations in the optical (Johnson and Hipparcos photometry converted to absolute fluxes and spectrophotometry by \citet{Adelman89}) and near-IR (2MASS fluxes) wavelength regions. The formulas of \citet{Lipski08} were used to convert the Adelman magnitudes to fluxes, the 2MASS observations were converted using formulas from \citet{Cohen03} and the calibration for Johnson photometry by \citet{Bessell98} was used. 

According to the literature, 49 Cam is not significantly reddened \citep[e.g][]{Adel82}, so reddening was not included in the calculations. The temperature was selected by evaluation of which model gave the best fit (by visual inspection) between the model flux distributions and the observational data. The best fit was obtained for $T_{\rm eff}=7800$~K. The corresponding comparison between the observed and theoretical SED is shown in Fig.~\ref{SED}. Following the same procedure, we examined a series of SEDs for model atmospheres with different values of $\log g$ (each varying by a step of 0.25) and found that $\log g=4.0$ gave the best fit to observations. An uncertainty of 0.25 is assigned based on the step between different $\log g$ values tested. Therefore, we adopted $T_{\rm eff}=7800$~K and $\log g=4.0$ for the subsequent analysis of 49 Cam.

The {\sc LLmodels} model atmosphere calculations took into account individual stellar abundances for about a dozen chemical species, including Fe-peak and several rare earth elements. These abundances were determined by fitting the {\sc Synmast} \citep{Kochukhov10b} theoretical spectra to observations of 49 Cam, based on an initial solar abundance model. The model atmosphere calculations also incorporated a modification of the line opacity due to Zeeman splitting and polarised radiative transfer \citep{Khan06}. For these calculations we assumed a magnetic field modulus of 3~kG \citep{Glag11}. 

An initial estimate of the projected rotational velocity $v\sin i$ was obtained by fitting a selection of spectral lines, typically in the 4500--5000 \AA\ wavelength region, with a synthetic spectrum. The projected rotational velocity was found to be $23 \pm 1$~\kms, in agreement with the value reported by \citet{Silvester12}.

As a by-product of matching theoretical SEDs to observations we derived an angular diameter, which can be converted to the stellar radius, $R=2.25\pm0.25R_\odot$, for the parallax $\pi=10.34\pm0.45$~mas \citep{vanLeeuwen07}. Our radius value agrees well with the previous estimate by \citet[][$2.11R_\odot$]{Kochukhov06}, but is marginally inconsistent with $R=1.9R_\odot$ given by \citet{Leone00}. The radius uncertainty is calculated based on the parallax uncertainty and the uncertainty from the derived angular diameter. 

Rotational phases of observations were calculated according to the ephemeris by \citet{Adelman97}:  ${\rm JD}=2441257.300 + 4\fd 28679\cdot{\rm E}$. The summary of derived stellar parameters for 49 Cam in this paper or adopted from the literature is given in Table~\ref{parameter-table}.

\begin{table}
\begin{center}
\caption{Parameters of 49 Cam.}
\begin{tabular}{ccc}
\hline
\hline  
Parameter & Value & Reference \\
\hline
\teff &  $7800 \pm 200$ K  &  this study  \\
\lgg & $4.00 \pm 0.25$ &  this study  \\ 
$P_{\rm rot}$ & 4.28679 days & \citet{Adelman97}  \\
$R$ & $2.25 \pm 0.25$ $R_\odot$ &  this study \\ 
\vs & $23.0 \pm 1.0 $ km\,s$^{-1}$ &  this study \\
$i$ & $120 \pm 10\degr$ &  this study \\
$\Theta$ & $40 \pm 5\degr$ & this study \\
\hline
\label{parameter-table}
\end{tabular}
\end{center}
\end{table}

\begin{table}
\begin{center}
\caption{Atomic lines used for the mapping of 49 Cam. The $\log gf$ values are adopted from the VALD3 database \citep{Ryabchikova15}.
\label{tbl:lines}}

\begin{tabular}{lcc}
\hline
\hline
Ion & Wavelength & $\log gf$ \\
&(\AA )  & \\
\hline
\multicolumn{3}{c}{Magnetic and abundance mapping}\\
\hline
 Ti\,{\sc ii}& 4394.059 & -1.770  \\   
                & 4395.031 & -0.540  \\   
                & 4779.985 & -1.260  \\   
Cr\,{\sc ii}  & 4588.199  & -0.627 \\
         & 4812.337 & -1.960   \\
         & 5407.604 & -2.151 \\
Fe\,{\sc ii}& 4620.513 & -3.240 \\
             & 4923.927 & -1.320 \\
         & 5018.440 & -1.220  \\
         & 5169.033 & -1.303  \\
Fe\,{\sc i}   & 6230.722 & -1.281 \\
Nd\,{\sc iii}& 5050.695 & -1.060  \\
                  & 5429.794 & -1.240  \\
                  & 5677.178 & -1.450 \\
                  & 5851.542 & -1.550 \\
 \hline                 
\multicolumn{3}{c}{Abundance mapping}\\       
\hline 
Ca\,{\sc i}  & 6122.217  & -0.316 \\
         & 6162.173  & -0.090   \\           
Sc\,{\sc ii}  & 5526.790  & 0.024 \\        
Ce\,{\sc ii}  & 4562.359  & 0.210 \\       
Pr\,{\sc iii}  & 7030.386  & -0.929 \\                 
Eu\,{\sc ii}  & 6437.631  & -0.602 \\     
                   & 6645.094 & -0.162 \\      
\hline
\label{line-list}
\end{tabular}
\end{center}
\end{table}

\section{Magnetic Mapping and Chemical Inversions}
\label{sect:methods}
To reconstruct the magnetic field of 49 Cam, we performed inversions of Stokes $IQUV$ profiles using {\sc Invers10} \citep{Piskunov02}. The magnetic Doppler imaging in this study follows very closely the methodology used in our previous mapping of the Ap stars $\alpha^2$\,CVn \citep{Silvester14a,Silvester14b} and HD 32633 \citep{Silvester15}. 

Within the inversion code the magnetic field topology is parametrised in terms of a general spherical harmonic expansion (up to the angular degree $\ell=10$ in this study), as described by \citet{Koch14}. To facilitate convergence to a global chi-square minimum a regularisation function is used. In the case of spherical harmonics the regulation function acts to limit the field structure to the simplest harmonic distribution that provides a satisfactory fit to the data.

The regularisation was chosen by a stepwise approach, with each successive step of decreasing levels of regularisation. Following the methodology described by \citet{Kochukhov17}, the final value of regularisation was chosen at the point in which the fit between the observations and the model was no longer significantly improved by decreasing regularisation.  Typically we choose a value of 2--5 for the ratio of total deviation to total regularisation. This approach was used both for the magnetic and abundance maps. 

The field reconstruction requires knowledge of several stellar properties.  The projected rotational velocity (\vs) and the inclination ($i$, the angle between the rotational axis and the line of sight) must be known. In addition, modelling of the Stokes $QU$ observations requires the knowledge of the azimuth angle ($\Theta$) of the stellar rotational axis. To verify our initial determination of $ v\sin i$ we performed an additional series of test inversions with different $v\sin i$ values, and confirmed that a $ v\sin i$ of 23 km\,s$^{-1}$ did result in the smallest deviation between the model spectra and the observations. 

To determine an inclination angle, we considered values reported for 49~Cam in the literature, such as 61\degr\ given by \citet{Stepien89} and  90\degr\ used by \citet{Leroy96}. We produced a grid of inversions with $i$ covering this range and selected the value that gave the smallest deviation between the model spectra and the observations.  As a result of this procedure, we initially found the best fitting inclination $i=60\degr$, which also agreed with the inclination angle that follows from the stellar $v\sin i$, radius and rotational period.  However for any given derived inclination angle, $i=180\degr-i$ is also a possible alternative solution. To check which of the two possible solutions gave the best fit, inversions were also performed for $i=180\degr-60\degr=120\degr$.  We found that $i=120\degr$ gave a slightly improved fit to the data and we therefore adopted that as our final value $i=120\degr \pm 10\degr$. It should be noted we do not find large differences between the final $i=60\degr$ and $i=120\degr$ maps, thus either option gives comparable results.
 
 As discussed by \citet{Kochukhov2004}, an uncertainty of 10\degr\ in the inclination angle is deemed to have little effect on the resulting maps.  A $\Theta$ angle of 20\degr\ was also determined on the basis of s similar grid inversion approach.

\begin{figure*}
\begin{center}
    \includegraphics[width=0.65\textwidth, angle=-90]{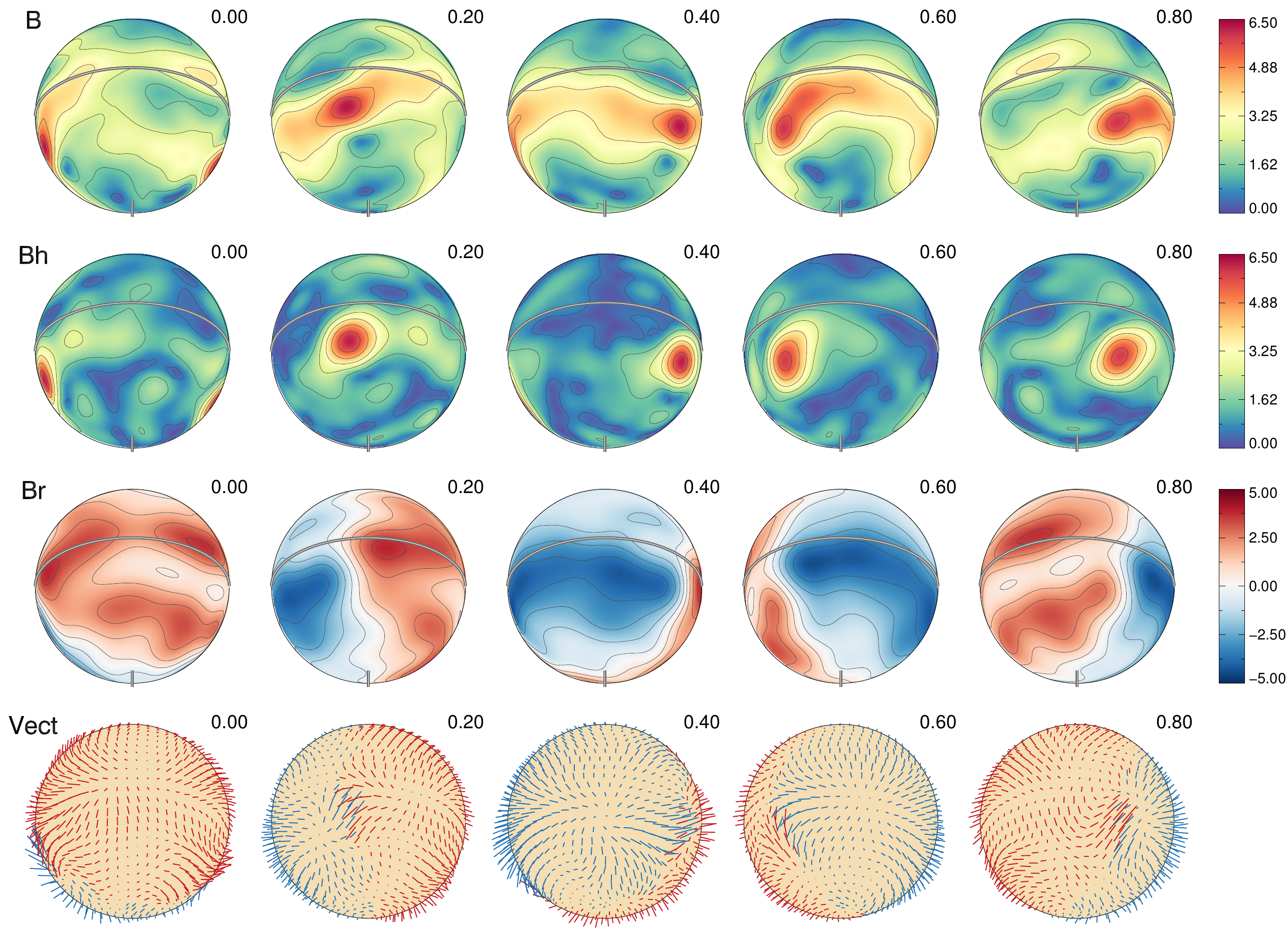}
  \caption{Surface magnetic field distribution of 49 Cam derived using Stokes $IQUV$ 
profiles. The spherical plots, displayed at the inclination angle of $ i = 120\degr$, show distributions of: the field modulus, horizontal field, the radial field, and the field orientation. Each column corresponds to a different phase of rotation (0.0, 0.2, 0.4, 0.6, and 0.8) and the colour bars on the right side indicate the magnetic field strength in kG. The contours over the top three maps are plotted with a 1.0 kG step.}
\label{maps-field}
\end{center}
\end{figure*}

\begin{figure*}
\begin{center}
       \includegraphics[width=0.65\textwidth, angle=-90]{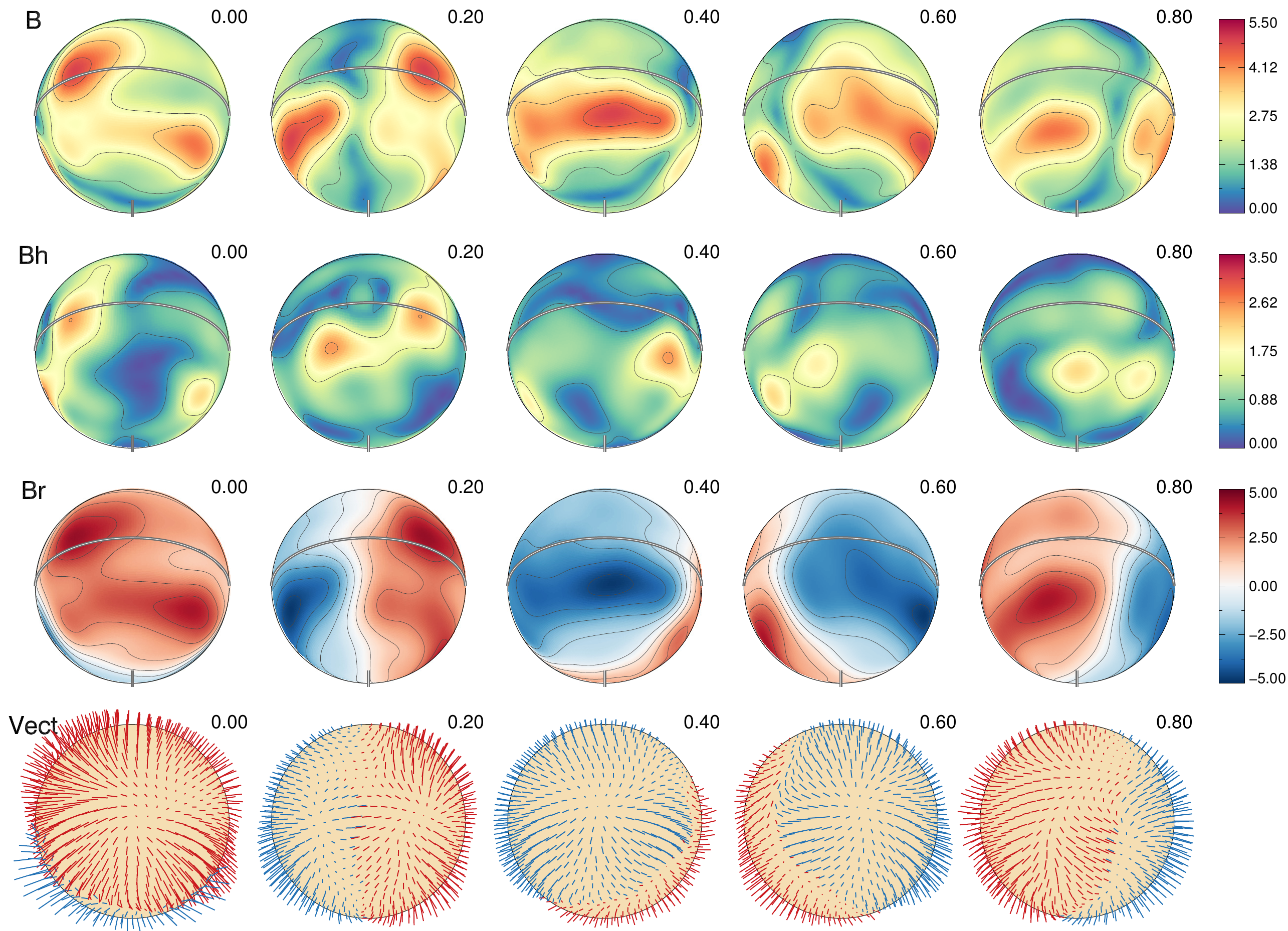}
  \caption{Same as Fig.~\ref{maps-field} but for the surface magnetic field distribution reconstructed using Stokes $IV$ spectra.}
\label{maps-field-iv}
\end{center}
\end{figure*}

\begin{figure*}
\begin{center}
  \includegraphics[width=0.99\textwidth]{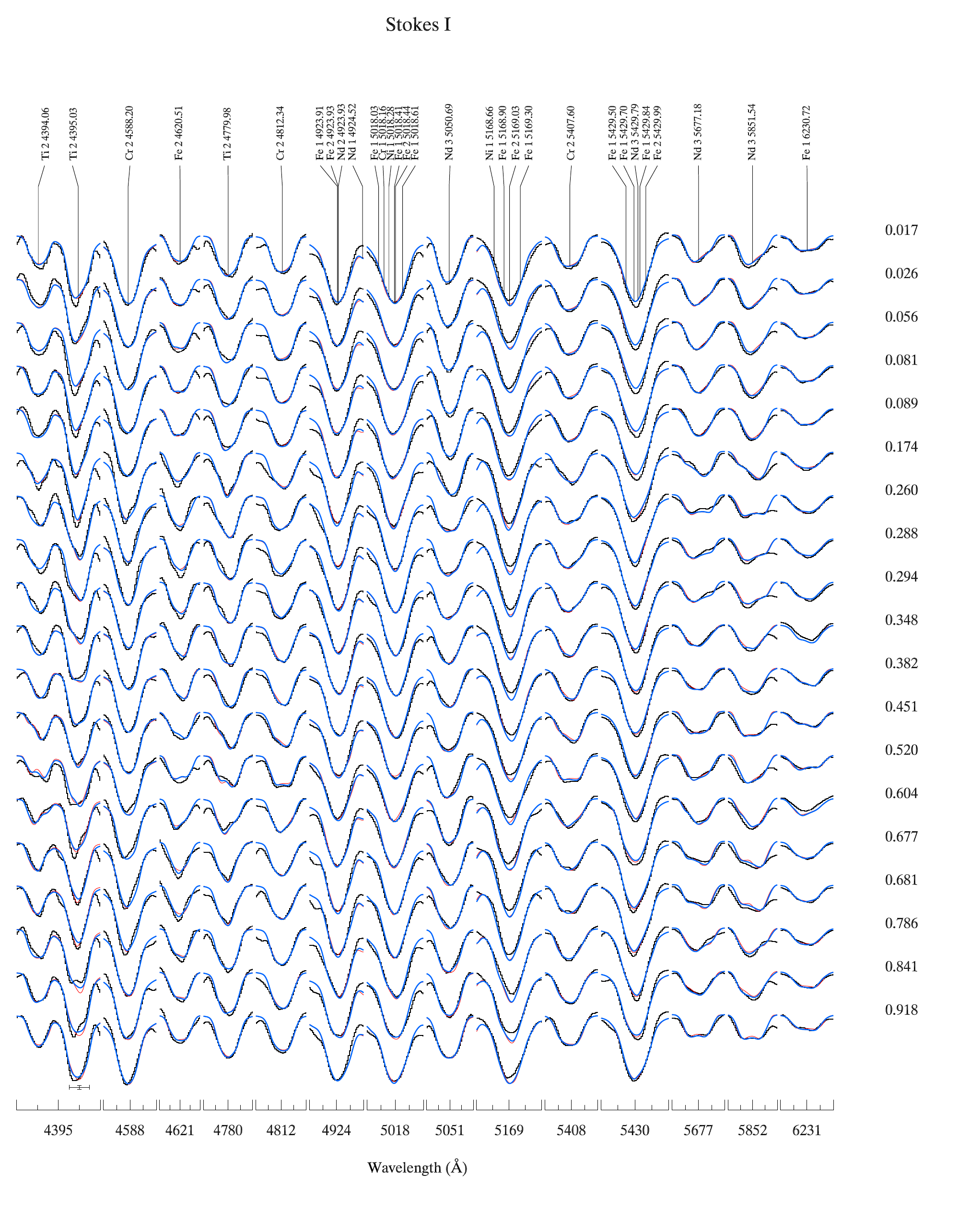}
   \caption{Stokes $I$ profiles for all lines used in the magnetic field mapping. Observed spectra (black histogram line) are compared to the synthetic profiles corresponding to the general spherical harmonic inversion (thick blue lines) and to the inversion limited to $\ell=2$ harmonic modes (thin red line). The spectra are offset vertically according to the rotational phase indicated to the right of the panel. The bars in the lower left corners indicate the vertical and horizontal scales (2 \% of the Stokes $I$ continuum intensity and 0.5 \AA).}
\label{Fit-I-Fld}
\end{center}
\end{figure*}

\begin{figure*}
\begin{center}
   \includegraphics[width=0.99\textwidth]{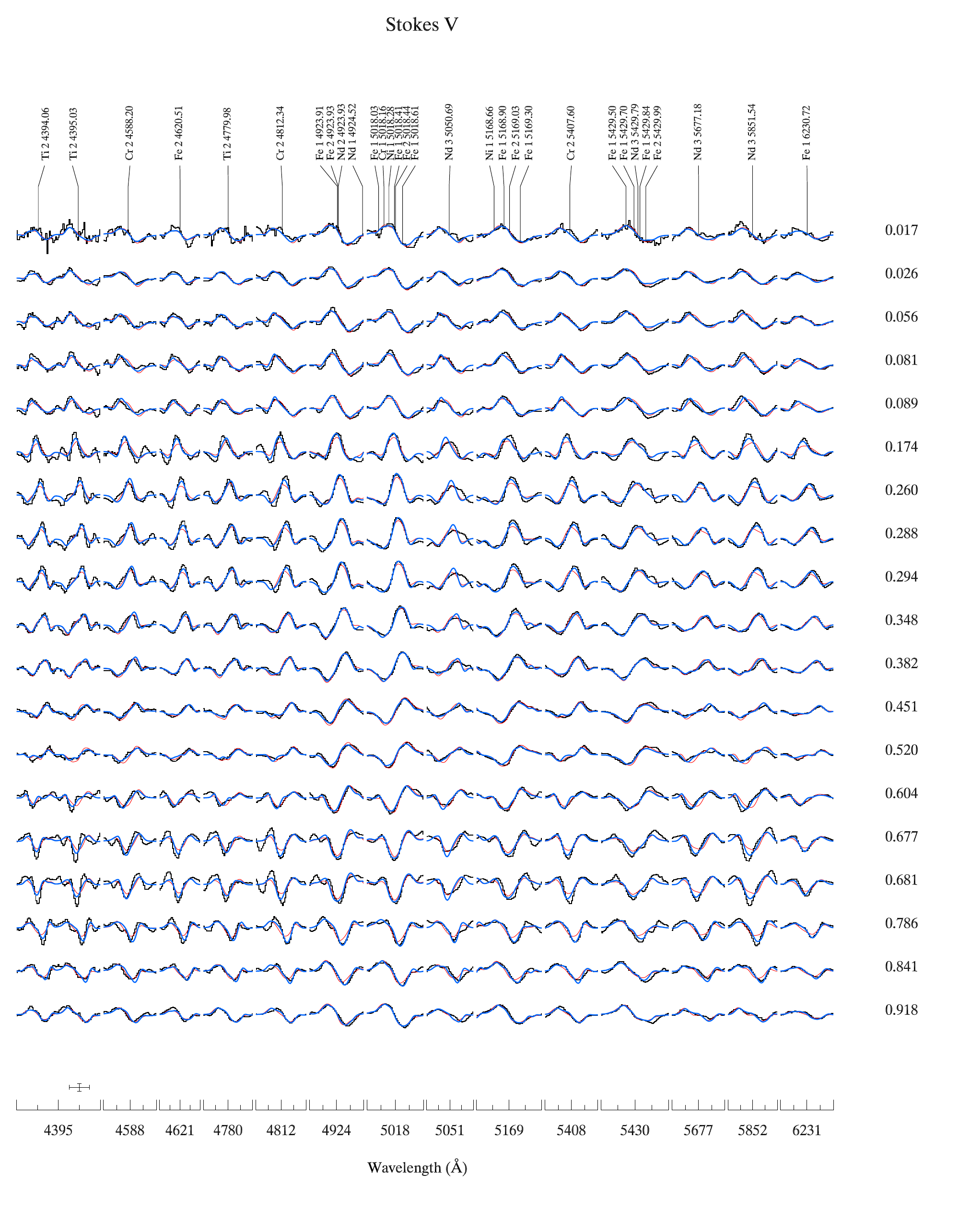}
   \caption{Same as Fig.~\ref{Fit-I-Fld} for the Stokes $V$ profiles.}
\label{Fit-V-Fld}
\end{center}
\end{figure*}

\begin{figure*}
\begin{center}
 \includegraphics[width=0.99\textwidth]{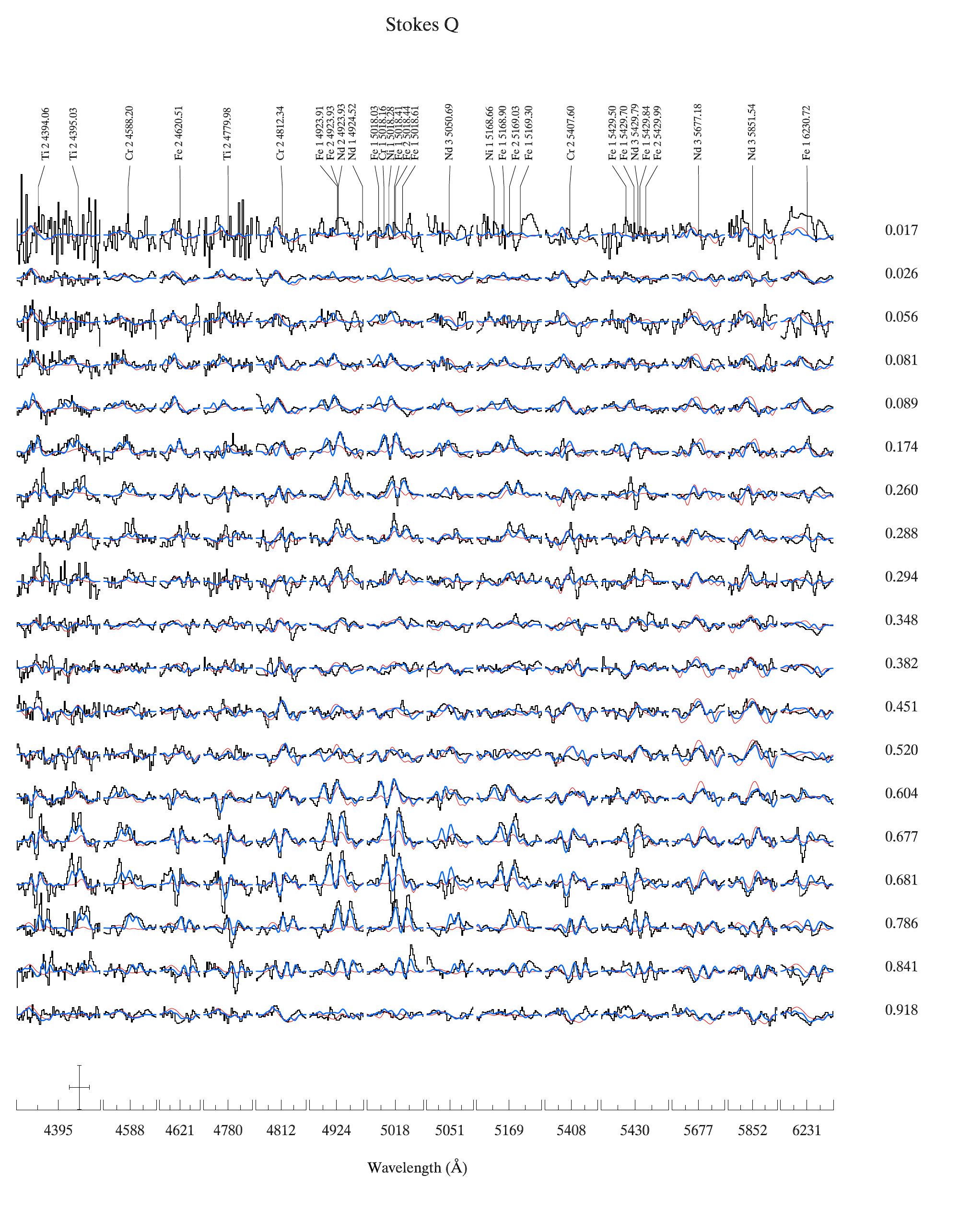}
   \caption{Same as Fig.~\ref{Fit-I-Fld} for the Stokes $Q$ profiles.}
\label{Fit-Q-Fld}
\end{center}
\end{figure*}

\begin{figure*}
\begin{center}
  \includegraphics[width=0.99\textwidth]{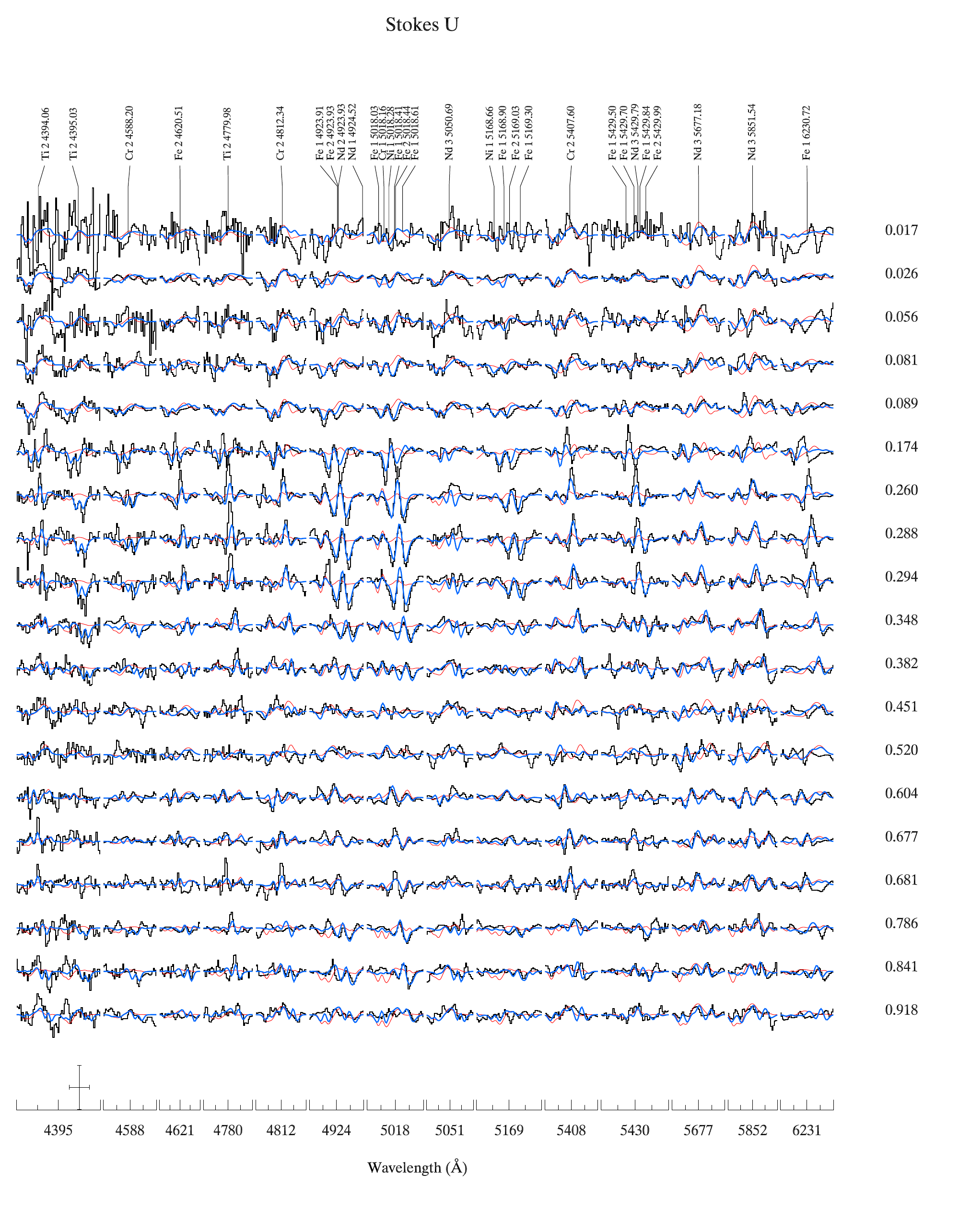}
   \caption{Same as Fig.~\ref{Fit-I-Fld} for the Stokes $U$ profiles.}
\label{Fit-U-Fld}
\end{center}
\end{figure*}

\section{Magnetic Field Structure}
\label{sect:mag}
The final magnetic field map for 49 Cam was derived using a set of Stokes $IQUV$ profiles for Ti, Cr, Fe and Nd lines.  These lines were selected on the basis that they contain no significant blends and that they exhibited strong polarisation signatures.  For the magnetic mapping all the lines selected were modelled simultaneously and the chemical abundance distribution maps for those elements were also derived simultaneously.  The relative weights of the Stokes $IQUV$ contributions are determined and assigned automatically by the code. 

Information on atomic parameters of these lines is given in Table.~\ref{line-list}. The resulting magnetic field map is presented in Fig. \ref{maps-field}. For comparison we also derived the magnetic field ignoring information in the Stokes $QU$ profiles. This magnetic map is shown in Fig.~\ref{maps-field-iv}. The fits between the model profiles and the observed profiles are illustrated in Figs.~\ref{Fit-I-Fld}, \ref{Fit-V-Fld}, \ref{Fit-Q-Fld} and \ref{Fit-U-Fld}. It should be noted that a limited number of Stokes $QU$ profiles are poorly fit at certain phases. However, these poorly fit profiles are limited to stronger lines and thus this suggests that this discrepancy is likely due to chemical stratification of these elements. This phenomenon is well know to occur in cooler Ap stars such as 49 Cam and is not taken into account in the inversion code.  There are some limited instances where overfits to the Stokes $QU$ profiles are seen (e.g for the phases of 0.677 and 0.681 in Nd 3 lines).  To get the best representation of the magnetic field, the field is derived using as many elements as possible, this will in some cases result in a small number of over-fitted or under-fitted regions.  This is an unavoidable consequence of using many lines,  it should be noted however that in the case of the overfit to these discrete Nd profiles, the overall contribution to the final magnetic field derived is negligible. 

\begin{figure}
\begin{center}
    \includegraphics[width=0.50\textwidth]{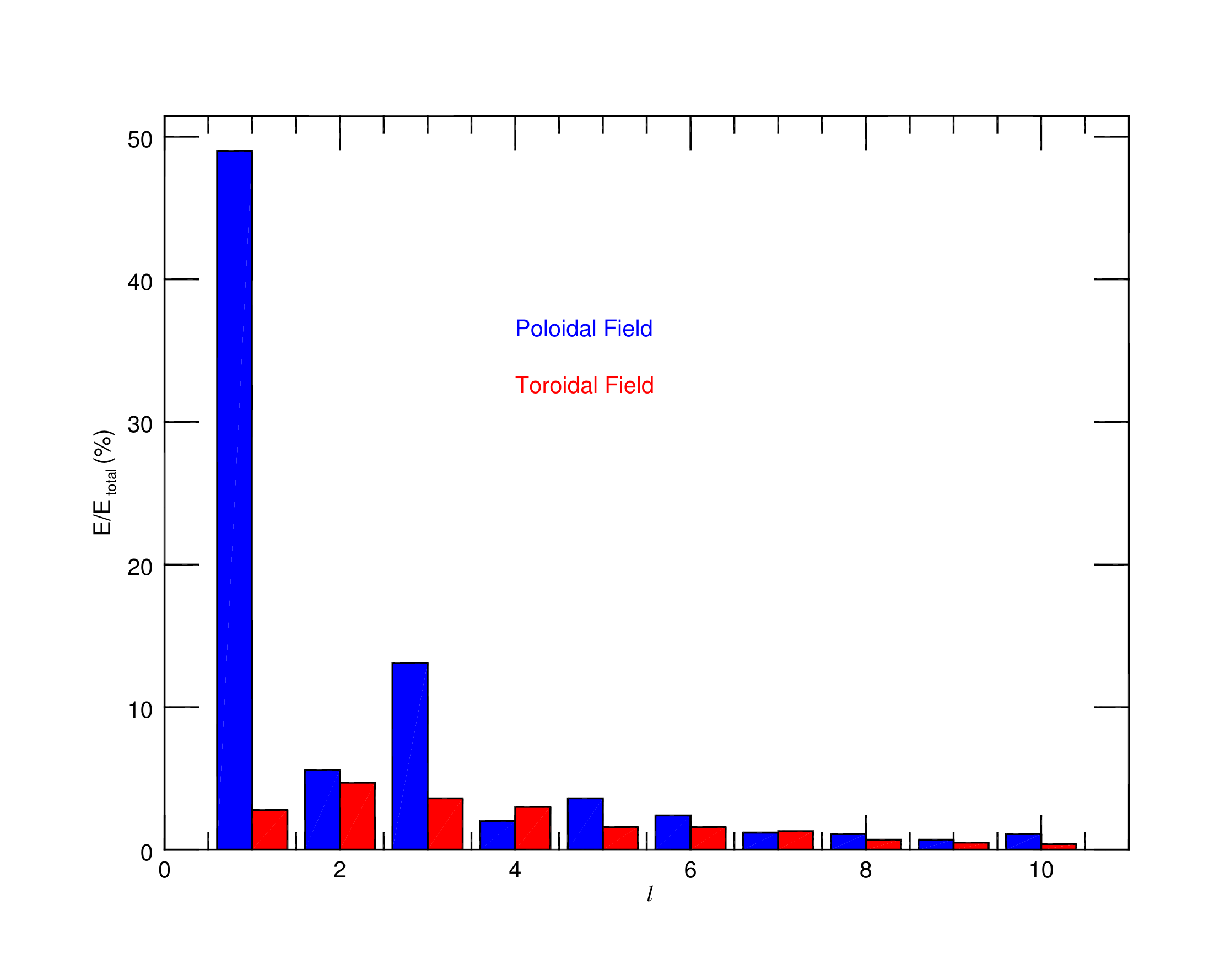}
   \caption{Distribution of the magnetic energy between different spherical harmonic modes in the MDI magnetic field map of 49 Cam. Bars of different colour represent the poloidal (blue/dark) and toroidal (red/light) magnetic components.}
\label{power-plot}
\end{center}
\end{figure}

The magnetic field of 49 Cam shows a relatively complex structure. In particular, the field topology of this star is considerably more complex than that of HD\,24712 \citep{Rusomarov15}, which is the only other cool Ap star mapped using line profiles in all four Stokes parameters. Contributions of different harmonic terms to the field topology are reported in Table~\ref{energies}. Considering this table and Fig. \ref{power-plot}, there is a significant contribution of the $\ell=1,2$ harmonic modes and large contributions of the $\ell=3,4$ modes.  This is reflected in the complex nature of the vector magnetic field map seen in Fig.~\ref{maps-field}, which shows two structures near the stellar equator (seen clearly at phases 0.20 and 0.80), these regions also correspond to the large horizontal field structures seen, for example at a phase of 0.20.

To test the reliability of magnetic maps and to verify the complexity of the field, inversions were performed with the maximum angular degree of spherical harmonics limited to $\ell=2$, thus forcing the field to remain simple in structure.  In the resulting inversions both Stokes $Q$ and $U$ profiles were extremely poorly fit, as can be seen from Figs.~\ref{Fit-I-Fld}, \ref{Fit-V-Fld}, \ref{Fit-Q-Fld} and \ref{Fit-U-Fld}. This strongly indicates that the magnetic field of 49 Cam cannot be described with a superposition of simple dipole and quadrupole components.

An important related test that we have conducted was an assessment of the contribution of information in the linear polarisation (Stokes $QU$) profiles to the magnetic inversion. To this end, we repeated inversions using only Stokes $I$ and $V$ observations. The resulting magnetic map is shown in Fig. \ref{maps-field-iv} and the distribution of energies for the $\ell$\,=\,1--4 modes is given in Table.~\ref{IVenergies}. The magnetic field topology inferred from modelling of the Stokes $IV$ data is somewhat simplified compared to the outcome of the four Stokes parameter inversion. In particular, we note that the horizontal field is much weaker due to the absence of linear polarisation information. However, we still find significant contributions of $\ell=2$ and 3 harmonic modes.

Compared to the harmonic magnetic energy spectrum of both $\alpha^2$~CVn \citep{Silvester14b} and HD\,32633 \citep{Silvester15}, the magnetic field of 49 Cam appears to be much more complex, with more energy in higher harmonic modes. In the case of these other stars, the harmonic energies were concentrated primarily to the first two modes with $\ell=1,2$, with higher $l$ contributions limited to few percent. Another cool Ap star ($T_{\rm eff}=9850$~K) which has also shown some higher order complexity is CS Vir \citep{Rusomarov16} with a total $l=3$ contribution of 8.9 \%, which is only slightly more than one-half the $l=3$ contribution of 49 Cam. Thus 49 Cam remains the Ap star with the largest $l=3$ contribution mapped to date.

\begin{table}
\begin{center}
\caption{Distribution of the poloidal ($E_{\rm pol}$) and toroidal ($E_{\rm tor}$) magnetic field energy over different harmonic modes for the best-fitting Stokes $IQUV$ MDI map of 49~Cam.}
\begin{tabular}{lccc}
\hline
\hline
$\ell$ & $E_{\rm pol}$ & $E_{\rm tor}$ & $E_{\rm tot}$ \\
 & (\%) & (\%) & (\%) \\
\hline
 1  &  49.0 &  2.8 &  51.8 \\
  2  &   5.6  &  4.7   & 10.3\\
  3   & 13.1  &  3.6  & 16.7\\
  4    & 2.1   & 3.0  & 4.9\\
  5    & 3.6   & 1.6   &  5.2\\
  6    & 2.4   & 1.6   & 4.0\\
  7    & 1.2   & 1.3   & 2.5\\
  8    & 1.1   & 0.7   & 1.8\\
  9    & 0.7   & 0.5   & 1.2\\
 10    & 1.1   & 0.4   & 1.5\\
 \hline
Total   & 79.7   & 20.3 & 100 \\
\hline
\label{energies}
\end{tabular}
\end{center}
\end{table}

\begin{table}
\begin{center}
\caption{Distribution of the poloidal ($E_{\rm pol}$) and toroidal ($E_{\rm tor}$) magnetic field energy truncated for the first four harmonics for the Stokes $IV$-only fit.}
\begin{tabular}{lccc}
\hline
\hline
$\ell$ & $E_{\rm pol}$ & $E_{\rm tor}$ & $E_{\rm tot}$ \\
 & (\%) & (\%) & (\%) \\
\hline
  1   & 74.7  &  1.9  & 76.6 \\
  2    & 5.6   & 1.8   & 7.3 \\
  3    & 6.5   & 2.2    & 8.7 \\
  4    & 1.1    & 1.1    & 2.2 \\
\hline
\label{IVenergies}
\end{tabular}
\end{center}
\end{table}

\begin{figure*}
\begin{center}
 \vspace{5 mm}
 \includegraphics[width=0.87\textwidth,angle=-90]{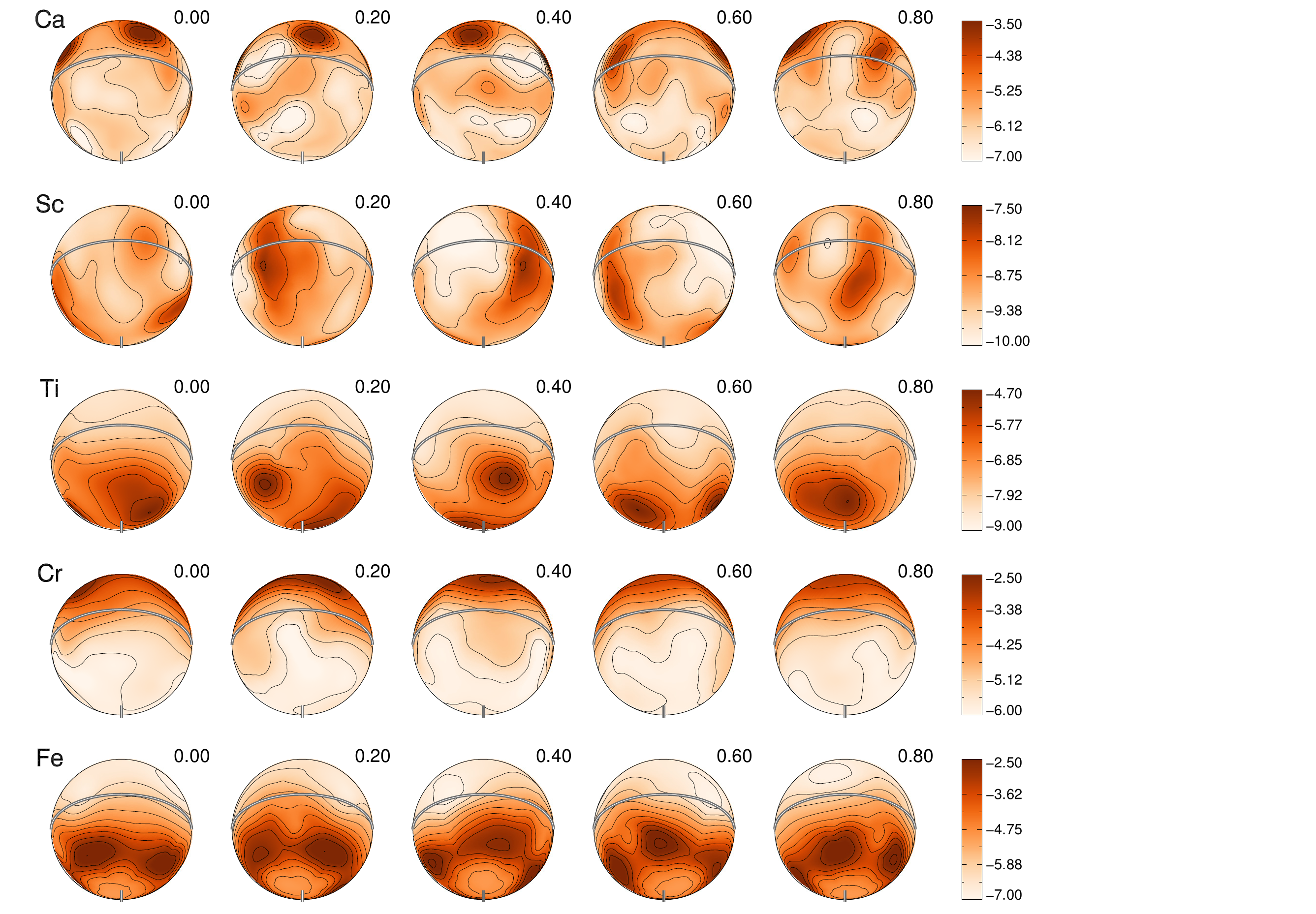}
   \includegraphics[width=0.202\textwidth,angle=90]{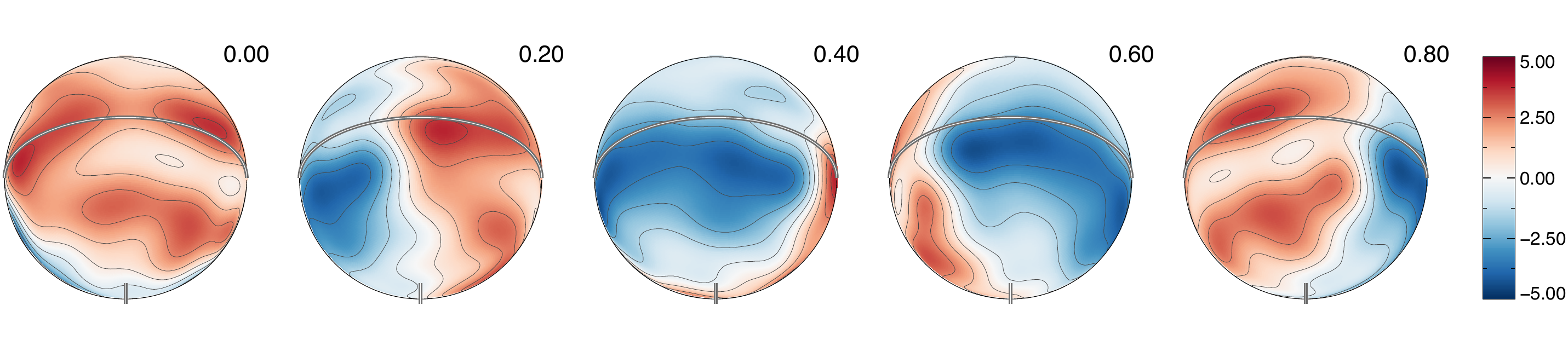}
    \caption{Chemical abundance distributions derived for Ca, Sc, Ti, Cr and Fe.  Each column corresponds to a different rotational phase (0.0, 0.2, 0.4, 0.6 and 0.8). The solid line shows the location of the stellar equator. The visible rotational pole is indicated by the short thick line. The colour bars on the right side indicate chemical abundance in the $\log N_{\rm el}/N_{\rm tot}$ units. The abundance contours are plotted with a 0.5 dex step. The radial magnetic field from the Stokes $IQUV$ is shown in the bottom row for comparison.}
\label{Maps-Abn1}
\end{center}
\end{figure*}

\begin{figure*}
\begin{center}
 \vspace{5 mm}
 \includegraphics[width=0.70\textwidth, angle=-90]{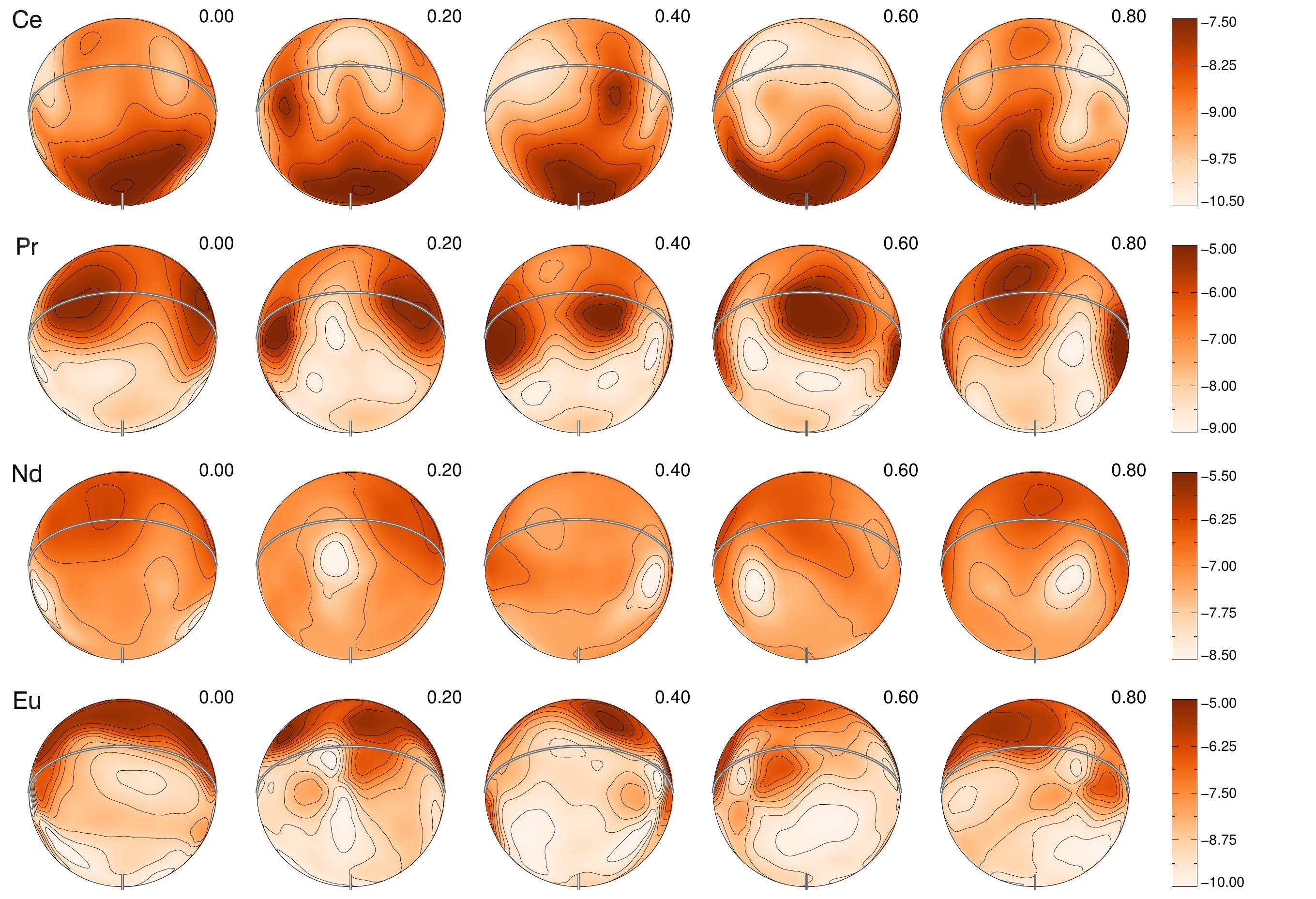}
   \includegraphics[width=0.202\textwidth,angle=90]{radial-field}
    \caption{Same as Fig.~\ref{Maps-Abn1} for the abundance distributions of Ce, Pr, Nd and Eu, with the radial magnetic field shown in the bottom row.   
}
\label{Maps-Abn2}
\end{center}
\end{figure*}

\section{Chemical Abundance Distributions}
\label{sect:abn}
We selected which chemical elements to map using the same methodology as implemented by \citet{Silvester14b} for $\alpha^2$~CVn and by \citet{Silvester15} for HD\,32633. Because of the lower effective temperature however and the different abundance pattern of 49 Cam, combined with its significant $v\sin i$, many spectral lines are heavily blended and the abundance maps are thus necessarily based on a limited set of essentially unblended lines. 

The final derived chemical abundance maps were produced using either inversions where the abundance distributions were derived simultaneously with the magnetic field using Stokes $IQUV$ or the abundance distributions were derived using Stokes $IV$ observations, with the magnetic field used as a fixed input, with this field being derived from the full Stokes $IQUV$ inversion.  For 49 Cam Ti, Cr, Fe and Nd were derived simultaneously with the magnetic field (and with Stokes $IQUV$) and Ca, Sc, Ce, Pr and Eu were derived with the fixed magnetic field (and from Stokes $IV$).  The list of lines used for abundance DI mapping is given in Table~\ref{line-list}.

For the final abundance spherical maps, an abundance scale was chosen such that extreme abundance value outliers were excluded from the final plot. The outliers normally comprised 2--12\% of all the surface elements and were selected with the use of a histogram plot, this was the same procedure as described by \citet{Silvester14b} and \citet{Silvester15}.

\subsection{Calcium and Scandium}
\begin{figure*}
\begin{center}
  \includegraphics[width=0.82\textwidth]{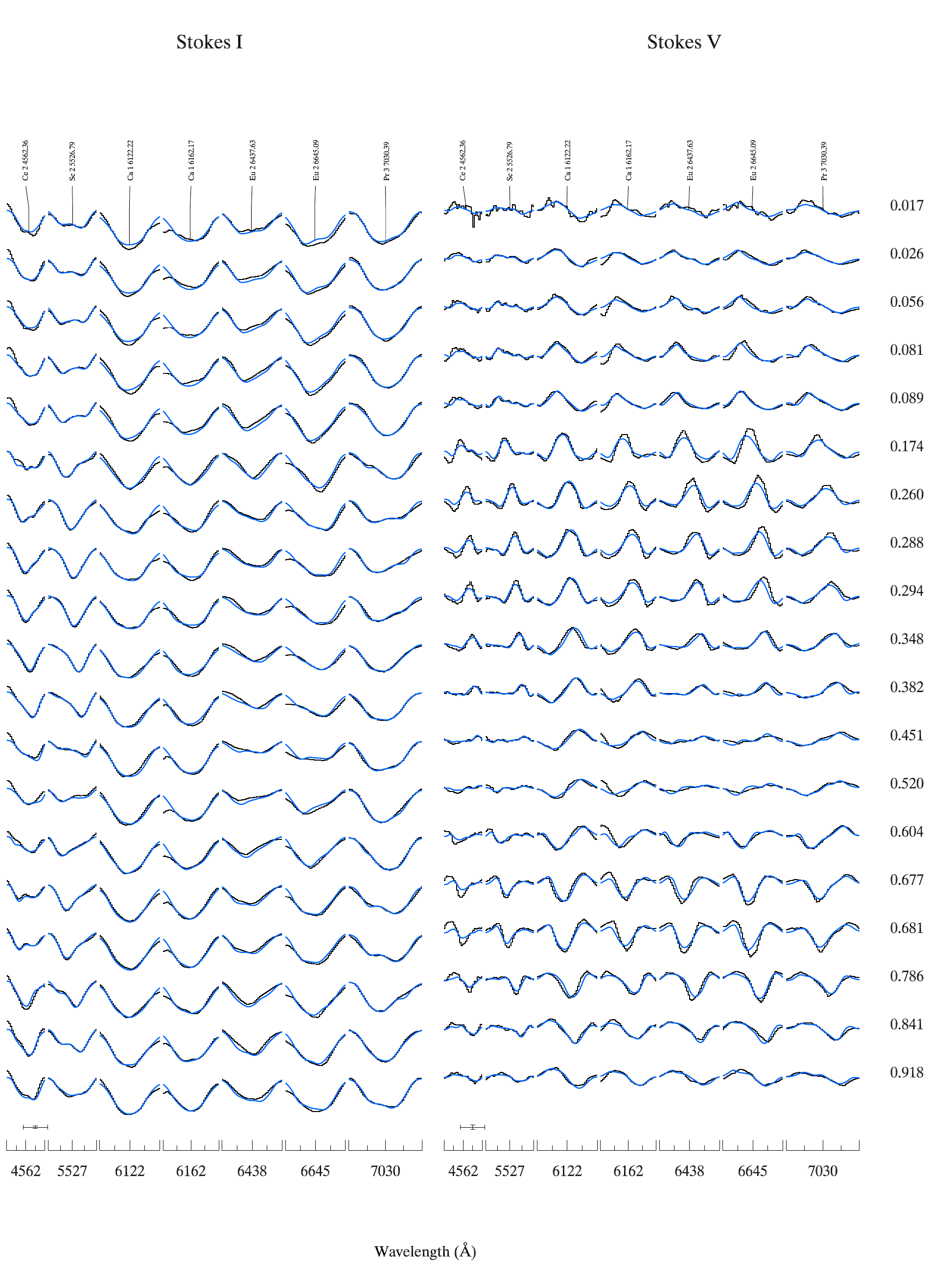}
\caption{Observed (black histogram line) and synthetic (solid blue line) Stokes $IV$ spectra of Ca, Sc, Ce, Pr, Eu lines used for abundance mapping of these elements.}
\label{Abn-Fits2}
\end{center}
\end{figure*}

The calcium abundance map was reconstructed using the Stokes $IV$ profiles of the Ca~{\sc i} $\lambda$ 6122 and 6162~\AA\ lines. The comparison of observations and the model profiles can be seen in Fig.~\ref{Abn-Fits2} and the resulting map is presented in Fig.~\ref{Maps-Abn1}. The calcium abundance ranges from $-3.5$ dex to $-7.0$ dex in the $\log N_{\rm el}/N_{\rm tot}$ scale (the solar value is $-5.70$). There are distinct areas of enhancement localised to spot-like regions close to the stellar equator.  The location of these distinct spots is difficult to clearly correlate to the magnetic field. 

The scandium abundance map was reconstructed using the Stokes $IV$ profile of the Sc~{\sc ii} $\lambda$ 5526~\AA\ line. The comparison of observations and the model profiles is shown in Fig.~\ref{Abn-Fits2} and the resulting map is presented in Fig.~\ref{Maps-Abn1}. The scandium abundance ranges from $-7.5$ dex to $-10.0$ dex  (the solar value is $-8.89$). Similar to calcium, scandium is located in distinct areas of enhancement, larger in extent than what is seen for calcium.  The enhancements are located predominantly near regions where the radial magnetic field component is weak and the horizontal component is strong. 

\subsection{Titanium, Chromium and Iron}

For titanium the abundance map was produced from the Stokes $IQUV$ profiles of the Ti~{\sc ii} $\lambda$ 4394, 4395 and 4779~\AA\ lines. The resulting comparison between the observations and the model profiles can be seen in Figs.~\ref{Fit-I-Fld}, \ref{Fit-Q-Fld}, \ref{Fit-U-Fld} and \ref{Fit-V-Fld} and the resulting  titanium distribution is given in Fig.~\ref{Maps-Abn1}. The titanium abundance has a range of $-4.7$ dex to $-9.0$ dex  (the solar value is $-7.09$). The titanium abundance is characterised by a pattern of enhancements located only in the bottom hemisphere, with a depleted area in the lower hemisphere.  The areas of enhancement do not appear to strongly correlate with the magnetic field.

A chromium map was reconstructed using the Stokes $IQUV$ profiles of the Cr~{\sc ii} $\lambda$ 4588, 4812 and 5407~\AA\ lines. The fit between observations and the model profiles is shown in Figs.~\ref{Fit-I-Fld}, \ref{Fit-Q-Fld}, \ref{Fit-U-Fld} and \ref{Fit-V-Fld}. The corresponding chromium distribution is displayed in Fig.~\ref{Maps-Abn1}. The chromium abundance has a range in value from $-2.5$ dex to $-6.0$ dex  (solar value is $-6.40$).  The largest chromium overabundance regions are located in the upper hemisphere.  No clear correlations between the magnetic field structure and the Cr abundance map are evident.

The iron map was reconstructed from the Stokes $IQUV$ profiles of the Fe~{\sc ii} $\lambda$ 4620, 4923, 5018, 5169~\AA\ and Fe~{\sc i} $\lambda$ 6230~\AA\ lines. The final fit between the observed and synthetic spectra is shown in Figs.~\ref{Fit-I-Fld}, \ref{Fit-Q-Fld}, \ref{Fit-U-Fld} and \ref{Fit-V-Fld}, with the resulting iron distribution given in Fig.~\ref{Maps-Abn1}.  The abundance range for iron is  $-2.5$ dex to $-7.0$ dex  (the solar value is $-4.54$). The iron abundance forms a very distinct ring of enhancement in the lower hemisphere around the area where the radial magnetic field is relatively high. This enhancement ring is broken at the location of the strong horizontal magnetic field patch near the magnetic equator.

\subsection{Cerium and Praseodymium} 

The cerium abundance map was reconstructed using the Stokes $IV$ profiles of the Ce~{\sc ii} $\lambda$ 4562~\AA\ line. The comparison of observations and the model profiles can be seen in Fig.~\ref{Abn-Fits2} and the resulting map is presented in Fig.~\ref{Maps-Abn2}. The cerium abundance ranges from $-7.5$ dex to $-10.5$ dex  (the solar value is $-10.46$). The cerium abundance distribution has large areas of enhancement and depletion. Comparing to the magnetic field, no clear correlations could be identified.

The praseodymium chemical map was reconstructed using the Stokes $IV$ profile of the Pr~{\sc iii} $\lambda$ 7030~\AA\ line. The comparison of observations and the model profiles can be seen in Fig.~\ref{Abn-Fits2} and the resulting map is presented in Fig.~\ref{Maps-Abn2}. The praseodymium abundance ranges from $-5.0$ dex to $-9.0$ dex  (the solar value is $-11.32$). The praseodymium abundance distribution has distinct spots with large areas of relative overabundance. These enhancements seem to be associated with the areas of strong radial magnetic field component. The areas of relative depletion seem to be located where the radial field is weak and the horizontal field is strong. 
  
\subsection{Neodymium and Europium} 

Neodymium was reconstructed using the Stokes $IQUV$ spectra of Nd~{\sc iii} $\lambda$ 5050, 5429, 5677 and 5851~\AA.  The comparison between observations and the model profiles is shown in Figs.~\ref{Fit-I-Fld}, \ref{Fit-Q-Fld}, \ref{Fit-U-Fld} and \ref{Fit-V-Fld}. The resulting map is shown in Fig. \ref{Maps-Abn2}. The neodymium abundance has a range of $-5.5$ dex to $-8.5$ dex  (the solar value is $-10.62$). The neodymium distribution is a less extreme version of the praseodymium distribution, with the largest enhancements also seen in areas where the positive radial component is relatively strong. The areas of lowest abundance seem to be located at the magnetic equator, where the horizontal field reaches maximum.

The europium abundance map was reconstructed using the Stokes $IV$ profiles of the Eu~{\sc ii} $\lambda$ 6437 and 6645~\AA\ lines. The comparison of observations and the model profiles can be seen in Fig.~\ref{Abn-Fits2} and the resulting map is presented in Fig.~\ref{Maps-Abn2}. The europium abundance ranges from $-5.0$ dex to $-10.0$ dex  (the solar value is $-11.52$). The europium distribution is similar to the praseodymium distribution, with areas of enhancement seen where the radial field is somewhat stronger and depleted areas near regions of weaker radial component and stronger horizontal field component. The europium lines used here have both hyperfine and isotopic structure, neither of which was included in our inversions. The same lines were used in Eu mapping for $\alpha^2$\,CVn \citep{Silvester14b} and it was verified that including their fine structure does not lead to an appreciable improvement in the fit quality or a change in the resulting abundance map.

\section{Conclusion and Discussion}
\label{sect:discuss}

We have carried out a magnetic Doppler imaging modelling of the full Stokes vector, high-resolution observations of the cool Ap star 49 Cam. This star is found to have a complex magnetic field structure combined with distinct abundance patterns. The magnetic field has significant contributions of the harmonic modes with $\ell$ up to 4, with a significant toroidal component. The necessity for such a complex field is supported by the fact that the observed Stokes $IQUV$ profiles cannot be reproduced by using a simpler field geometry and also by the relative complexity of the field structure inferred from the Stokes $IV$ inversions. The magnetic field of 49 Cam is significantly more complex than the fields seen of $\alpha^2$\,CVn and HD 32633 \citep{Silvester14b,Silvester15} mapped in four Stokes parameters using the comparable ESPaDOnS/Narval dataset and is dramatically more structured than the field of the cool Ap star HD\,24712 studied with the HARPSpol Stokes $IQUV$ data \citep{Rusomarov15}. Our conclusions are qualitatively consistent with those of  \citep{Leroy94} , who used broadband linear polarisation measurements to model the magnetic field of 49 Cam, concluding that 49 Cam had a magnetic structure which departs noticeably from the customary dipole configuration. 

Being a cool Ap star, 49 Cam thus contradicts the marginal trend which has been seen in previous studies, in which simpler magnetic field configurations have been found in cooler Ap stars and complex fields have been typically limited to hotter stars. This suggests that stellar effective temperature may not be correlated with the magnetic field complexity in any meaningful way. It is still possible that the field complexity correlates with the stellar age. However, all four Stokes-parameter MDI targets are field stars for which age determination is notoriously uncertain \citep{Landstreet07}.

We find no clear correlations between the abundance distributions of Ca, Ti, Cr and Ce and the magnetic field structure. For the remaining chemical elements we find two distinct patterns. For Fe, Pr, Nd and Eu we find that the areas of relative overabundance are localised to the surface zones where the radial magnetic field is relatively strong and the areas of relative depletion are located where the horizontal magnetic component is strongest. A reverse trend is seen for Sc, with the areas of relative enhancements located where the horizontal field is strong.

Comparing these results to the abundance maps obtained in the studies of $\alpha^2$\,CVn and HD 32633 \citep{Silvester14b,Silvester15}, there is a somewhat limited overlap of the elements that were mapped in all three stars and that also show correlations with the magnetic field structure. In the case of $\alpha^2$\,CVn, the elements Fe, Pr, Nd and Eu show correlations with the magnetic field and in fact these correlations agree with those found for 49 Cam, with these elements being relatively overabundant in the areas of higher radial magnetic field and depleted at the areas near the magnetic equator or stronger horizontal field component.  In the case of HD 32633, only the elements Fe and Nd show clear correlations with the field. For Nd we see similar behaviour with an enhancement in the regions of stronger radial magnetic field component. In the case of iron in HD 32633 we find the opposite behaviour, with the areas of enhancement located near the magnetic equator. 

Considering our abundance mapping results in the light of theoretical radiative diffusion predictions by \citet{Alecian15}, there is only one element, iron, for which we find clear correlations with the magnetic field structure and that has been studied in this theoretical work. The theory predicts that the abundance enhancement must appear as a ring-like structure at the magnetic equator. Whilst in our maps of 49 Cam we find iron is indeed distributed in a ring-like structure, instead of being located on the magnetic equator as predicted by theory, we find the ring to be located in the areas of strong radial magnetic field, and this ring is broken by the magnetic equator. On the other hand, our observations do support the prediction by \citet{Michaud81}, who suggested iron to be enhanced where the field is vertical. 

It is worth noting that the Ap stars mapped recently have shown a variety of magnetic field substructure, with levels of complexity differing between stars. Perhaps with Ap/Bp stars showing this magnetic diversity (at least in the small scale magnetic field or in the higher order harmonic modes), the comparison of observed abundance structures to theoretical models based on simplified models of the magnetic field is not advisable. 

We suggest that, when assessing the agreement between the theoretically predicted and observed abundance structures, theoretical modelling should be done on star by star basis, taking into account the individual, often complex stellar magnetic field topologies. Just as early models of the magnetic field geometry of Ap/Bp stars assumed a simple dipolar structure, we now understand through advances in instrumentation and modelling techniques how complex these magnetic fields actually are.

\section*{Acknowledgments} 
O.K. acknowledges financial support from the Knut and Alice Wallenberg Foundation, the Swedish Research Council, and the Swedish National Space Board. GAW acknowledges Discovery Grant support from the Natural Sciences and Engineering Research Council (NSERC) of Canada. This work was based on observations obtained at the Canada-France-Hawaii Telescope (CFHT) which is operated by the National Research Council of Canada, the Institut National des Sciences de l'Univers of the Centre National de la Recherche Scientifique of France,  and the University of Hawaii.  Also based on observations obtained at the Bernard Lyot Telescope (TBL, Pic du Midi, France) of the Midi-Pyr\'en\'ees Observatory,  which is operated by the Institut National des Sciences de l'Univers of the Centre National de la Recherche Scientifique of France. 

\bibliographystyle{mnras}
\bibliography{astro_ref_v1}


\label{lastpage}

\end{document}